\documentclass{egpubl}
\usepackage[T1]{fontenc}
\usepackage{dfadobe}  

\usepackage{cite}  
\BibtexOrBiblatex
\electronicVersion
\PrintedOrElectronic
\ifpdf \usepackage[pdftex]{graphicx} \pdfcompresslevel=9
\else \usepackage[dvips]{graphicx} \fi

\usepackage{egweblnk} 

\usepackage{times}
\usepackage{graphicx}
\usepackage{amsmath}
\usepackage{amssymb}
\usepackage{array}
\usepackage[toc,page]{appendix}
\usepackage{soul}
\usepackage[export]{adjustbox}

\title{Image Morphing with Perceptual Constraints and STN Alignment}

\author[N. Fish, R. Zhang, L. Perry, D. Cohen-Or, E. Shechtman \& C. Barnes]
{\parbox{\textwidth}{\centering N. Fish$^{1}$, R. Zhang$^{2}$, L. Perry$^{1}$, D. Cohen-Or$^{1}$, E. Shechtman$^{2}$ and C. Barnes$^{2}$}
        \\
{\parbox{\textwidth}{\centering $^1$Tel Aviv University, Israel \qquad 
         $^2$Adobe Research, USA
       }
}
}

\begin{document}

\newcommand{\teaserfig}{1.4}

\teaser{
 \centering
  \includegraphics[width=16cm]{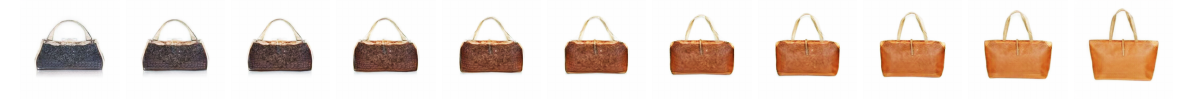}
 
  \caption{A morphing sequence generated by our approach.}
\label{fig:teaser}
}

\maketitle

\begin{abstract}
In image morphing, a sequence of plausible frames are synthesized and composited together to form a smooth transformation between given instances.
Intermediates must remain faithful to the input, stand on their own as members of the set, and maintain a well-paced visual transition from one to the next. In this paper, we propose a conditional GAN morphing framework operating on a pair of input images. The network is trained to synthesize frames corresponding to temporal samples along the transformation,
and learns a proper shape prior that enhances the plausibility of intermediate frames.
While individual frame plausibility is boosted by the adversarial setup, a special training protocol producing sequences of frames, combined with a perceptual similarity loss, promote smooth transformation over time. Explicit stating of correspondences is replaced with a grid-based freeform deformation spatial transformer that predicts the geometric warp between the inputs, 
instituting
the smooth geometric effect by bringing the shapes into an initial alignment. We provide comparisons to classic as well as latent space morphing techniques, and demonstrate that, given a set of images for self-supervision, our network learns to generate visually pleasing morphing effects featuring believable in-betweens, with robustness to changes in shape and texture, requiring no correspondence annotation.

\ccsdesc[500]{Computing methodologies~Image processing}
\ccsdesc[300]{Computing methodologies~Neural networks}

\printccsdesc

\end{abstract}

\keywords{image morphing, generative adversarial networks, spatial transformers, perceptual similarity}


\section{Introduction}

Morphing is the process of transformation between states of appearance, and may involve operations ranging from basic translation and rotation, to changes in color and texture, and, perhaps most iconically, shape shifting.
In the era of big data and deep learning, the ability to morph between objects could have an impact beyond the generation of the visual effect itself.
For instance, synthesized intermediate frames depicting novel variations of input objects, may be added to existing datasets for densification and enrichment.

Traditional morphing techniques rely on correspondences between relevant features across the participating instances, to drive an operation of warp and cross-dissolve \cite{beier1992feature}. These methods are mostly invariant to the semantics of the underlying objects and are therefore prone to artifacts such as ghosting and implausible intermediates. Correspondence points are normally user-provided, or are automatically computed assuming sufficient similarity. Recently, an abundance of available data has given rise to its utilization as guidance proxies for extraction of short or smooth paths between the two endpoints \cite{averbuch2016smooth}, thus providing more plausible in-betweens.

In this paper, we aim to further tap into the data-driven morphing paradigm, and leverage the power of deep neural networks to learn a shape prior befitting a given source dataset, catering to the task of image morphing. 
Specifically, we employ a generative adversarial network (GAN) \cite{goodfellow2014generative} combined with a spatial transformer \cite{jaderberg2015spatial} for shape alignment, for mitigation of the challenges associated with morphing.
GANs are known for displaying impressive generative capabilities by their capacity to learn and model a given distribution, a particularly lucrative attribute for a task for which realism and plausibility is crucial. Accordingly, we opt to design a GAN framework to learn the space of natural images of a given class so that intermediate frames appear to be realistic, and to enforce sufficient similarity between sufficiently close frames, to maintain smoothness of transformation.

We present a conditional GAN framework trained to generate sequences of transformations between two or more inputs, and further integrate it with a grid-based freeform spatial transformer network to alleviate large discrepancies in shape. The generated output sequences are constrained by a perceptual loss, culminating in an end-to-end solution that encourages transitions that are both plausible and smooth, with a gradual and realistic change in shape and texture.

The result is a trained generator specializing in a certain family of shapes, that, given a pair of inputs and a desired point in time, outputs the appropriate in-between frame. 
A full morphing effect can then be synthesized by requisitioning a reasonably dense sequence of frames, which yield a smooth transformation (see Figure \ref{fig:teaser}). 

During training, each sampled set of inputs is first processed by a spatial transformer network, which computes an alignment allowing a feature-based warp operation to map each input to the other. 
Next, our conditional generator processes the warped inputs, and outputs a sequence of frames, each corresponding to a given point in time.
A reconstruction loss encourages the two endpoint frames to match the inputs. Meanwhile, a GAN loss pushes the generated frames towards the natural image manifold of the training set. Finally, a perceptual transition loss \cite{zhang2018unreasonable} constrains the transformation over time to be smooth and gradual.

We demonstrate the competence of our generator and its ability to produce visually pleasing morphing effects with smooth transitions and plausible in-betweens, on different sets of objects, both real and computer rendered. We conduct a thorough ablation study to examine the individual contributions of our design components, and perform comparisons to traditional morphing, as well as GAN-based latent space interpolation. We show that our framework, uniting the GAN paradigm with shape alignment and perceptually constrained transitions, provides a solution that is robust to significant changes in shape, a challenging setup that commonly induces ghosting artifacts in morphs.


\section{Related Work}

\textbf{Classic morphing.}
Pioneering morphing techniques combine correspondence-driven bidirectional warping with blending operations to generate a sequence of images depicting a transformation between the entities in play. The approach by Beier and Neely \cite{beier1992feature} leverages user-defined line segments to establish corresponding feature points that are used to distort each endpoint towards the other, and proceeds to apply a cross-dissolve operation on respective pairs of warped images to obtain a transformation sequence.
More recently, Liao et al. \cite{liao2014automating} automatically extract correspondences for morphs by performing an optimization of a term similar to structural image similarity~\cite{wang2004image} on a halfway domain. Averbuch-Elor et al. \cite{averbuch2016smooth} adopt a data-driven approach where a collection of images from a specific class of objects is used to locate smooth sequences of images. A morphing effect is then generated from source to target via in-betweens that are smoothed with a global similarity transform. 
In deep image analogies \cite{liao2017visual}, deep features are leveraged for bidirectional correspondences supporting bidirectional attribute transfer for synthesis of style and content hybrids.
Similarly, Aberman~et~al.~\cite{aberman2018neural} focus on cross-domain correspondences extracted using a coarse-to-fine search of mutual nearest neighbor features, and show that this can produce cross-domain morphs. 
Shechtman~et~al. \cite{shechtman2010regenerative} introduced an alternative way to morph between different images using patch-based synthesis that did not rely on correspondences and cross-blending, and Darabi~et~al. \cite{darabi2012image} extended it by allowing patches to rotate and scale. While these methods produced nice transitions that look different than the traditional warp+blend effect, the method is limited to patches drawn from the two sources and does not produce new content.

\textbf{Deep interpolations.}
Neural networks have been previously trained to synthesize novel views of objects using interpolation. Given two images of the same object from two different viewpoints as input, Ji et al. \cite{ji2017deep} generate a new image of the object from an in-between viewpoint. The images are first brought into horizontal alignment, and are then processed by an encoder-decoder network that predicts dense correspondences used to compute an interpolated view.
General image interpolations are commonly demonstrated within the VAE and GAN realms. A notable by-product of a trained GAN is its rich latent embedding space that facilitates linear interpolation between data points. Such interpolations drive a generation of morphing sequences, by producing a series of interpolated latent vectors that are decoded to images that appear to smoothly morph from source to target \cite{brock2018large,berthelot2017began,karras2017progressive,donahue2019large}.
To perform interpolation between existing instances, one must obtain their corresponding latent codes in order to compute interpolated vectors and their decoded images. This is commonly accomplished with an optimization process that starts from a random code, which is updated to minimize a loss such as $L2$ on the desired image~\cite{webster2019detecting}. However, in practice, the learned manifold may not be able to reconstruct any given test set image, and obtaining the corresponding code to a given image may also be challenging. Solutions that combine an encoder mapping existing instances to the learned space, such as VAE-GAN \cite{larsen2015autoencoding} and CVAE-GAN \cite{bao2017cvae}, which is trained simultaneously, and iGAN \cite{zhu2016generative}, which is trained successively, alleviate this difficulty, but the crux of the problem remains, particularly when one seeks to map more unique entities.


\section{Method}

Our system combines several key components that together provide a robust solution for morphing effect generation.
We henceforth present these components and address the manner in which they cater to the three requirements, namely, frame realism with respect to the training set, smooth transitions, and input fidelity at the endpoints.

\subsection{Basic setup}

We use a convolutional GAN approach \cite{goodfellow2014generative,radford2015unsupervised} for our morphing. 
GANs have been demonstrated to perform highly sophisticated modeling of image training data~\cite{brock2018large}. This characteristic is appealing for our endeavor, as we seek to create sequences of transformation between entities belonging to a specific family of objects, \emph{i.e.}, our target distribution. 
Therefore, a GAN loss can help fulfill our first requirement of realism. In our implementation, we combine the Least-Squares GAN loss \cite{mao2017least} with two discriminators: a local PatchGAN \cite{li2016precomputed} discriminator and a global discriminator. We denote by $\mathcal{L}_D$ and $\mathcal{L}_G$ the GAN losses used to train $D$ and $G$ respectively, each by minimization of the corresponding sum:
\begin{equation}
    \mathcal{L}_D = {\mathcal{L}_{D}}_{\text{local}}^{\text{real}} + {\mathcal{L}_{D}}_{\text{global}}^{\text{real}} + {\mathcal{L}_{D}}_{\text{local}}^{\text{fake}} + {\mathcal{L}_{D}}_{\text{global}}^{\text{fake}}
\end{equation}

\begin{equation}
\label{eq:lossdg}
    \mathcal{L}_G = {\mathcal{L}^*_{G}}_{\text{local}}^{\text{fake}} + {\mathcal{L}^*_{G}}_{\text{global}}^{\text{fake}}
\end{equation}

The asterisk in Equation \ref{eq:lossdg} indicates the inversion of labels when $D$ is used to evaluate $\mathcal{L}_G$. 

Common image morphing operates on existing instances given as input, thus, accordingly, we opt for a special type of GAN known as the conditional GAN \cite{mirza2014conditional,isola2017image,zhu2017unpaired, kim2017learning}, whose output is directly influenced by one or more signals given as input.  
In our case, those signals include the two input images that are to be morphed, $I_A, I_B$, as well as a scalar $t$ specifying the desired time sample of the output in-between frame. We note that this could also be generalized to an arbitrary number $k \ge 2$ of input images to be morphed, along with a vector of interpolation parameters with $L_1$ norm of unity. Our conditional GAN consists of an encoder followed by a generator.

Our second requirement is smoothness of transitions. This is dealt with by a combination of a special training protocol and a suitable loss component. To better control and guide the generation to comply with our aim, at training time, for each input pair $I_A, I_B$, we generate a sequence of frames of length $k$. Each of these frames correspond to a predetermined time sample, and are uniformly sampled on the unit interval $[0, 1]$. This approach allows us to apply a loss component, $\mathcal{L}_T$, designed to constrain the similarity between frames, and encourage smooth transitions. More specifically, we make use of a pretrained neural network (VGG-16 \cite{simonyan2014very}) to obtain deep features of generated frames upon which perceptual similarity (PS) is computed \cite{zhang2018unreasonable}. As a frame-of-reference, we compute input pair PS: $\text{PS}_{4,5}(I_A, I_B) = \text{MSE}(\text{VGG}_{4,5}(I_A),  \text{VGG}_{4,5}(I_B))$, where $\text{VGG}_{4,5}(I_A)$ are all VGG features of $I_A$ extracted from layer groups 4 and 5 (out of 5). Using that, we define $\mathcal{L}_T$ as:

\begin{equation}
    \mathcal{L}_T = \max_{i=2..k}\{ \|\text{PS}_{4,5}(I_{i-1}, I_i) - (t_i - t_{i-1}) \cdot \text{PS}_{4,5}(I_A, I_B)\|^2\}
\label{eq:local_ps}
\end{equation}

That is, we constrain each frame to be a certain distance, in semantic feature space, from its preceding frame. This distance should ideally match the feature distance $PS_{4,5}(I_A, I_B)$ between the input images, after rescaling by the time between adjacent frames $t_i - t_{i-1}$.

The final component in our basic setup is a reconstruction loss, which encourages the endpoint frames in the sequence to match the inputs:

\begin{equation}
    \mathcal{L}_R = \text{MSE}(I_1,I_A) + \text{MSE}(I_k,I_B)
\end{equation}

\subsection{Alignment}
\label{sec:alignment}
The characteristic locality of convolutional networks is a known hindrance in situations where changes in shape are required. 
To support a wide range of inputs of varying shapes, we recognize the need for higher-level, semantic information to establish the relationship between the inputs, much like classic morphing techniques that rely on correspondences between points and features to drive a warping operation. Manually collecting correspondence points between instances in large datasets such as ours is intractable. Although it is possible to incorporate an automatic correspondence computation \cite{aberman2018neural}, we opt for an integrated end-to-end solution which is both computationally faster, and as we show later, can be more robust in cases where there are significant differences in shape.

\begin{figure}
\includegraphics[width=8.5cm]{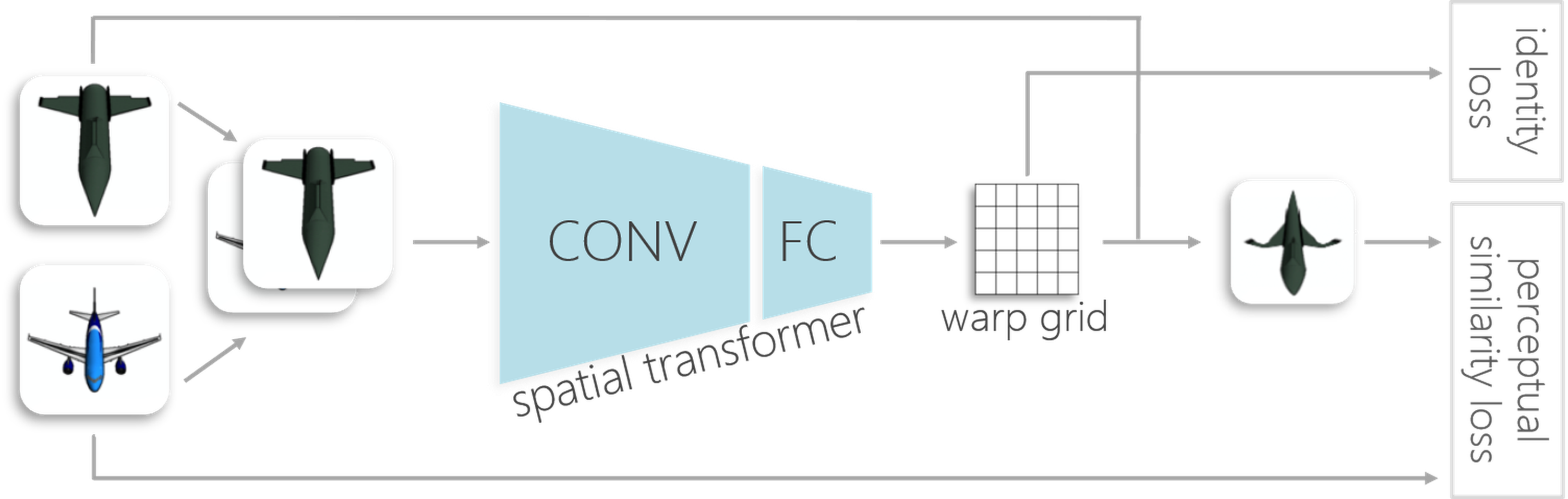}
\caption{Grid-based freeform spatial transformer. The two inputs are concatenated and processed by the network which outputs a 5x5 grid aligning the first to the second. The deformed first image is compared against the second image using perceptual similarity. The grid is compared to the identity grid for regularization.}
\label{fig:stn_arch}
\end{figure}

A spatial transformer network (STN) \cite{jaderberg2015spatial} is a component that can be added to a neural network as a means to learn and apply transformations to the data to assist the main learning task. In our setting, we seek to compute an alignment between the inputs, and apply it onto them to be given to the generator for further processing. For greater flexibility and range of deformation, we add a spatial transformer component that computes a grid-based freeform deformation warp field \cite{hanocka2018alignet}. This component is placed before the encoder-generator component of our main network, and is composed of two convolutional blocks and a fully-connected block 
predicting the warp grid, whose size is a parameter set to 5x5 in our experiments. 
The inputs $I_A, I_B$ are concatenated channel-wise before passing through this component, which outputs two grids (for $x$ and $y$) indicating the warp from $I_A$ to $I_B$ - $\mathcal{W}_{AB}$. Likewise, $\mathcal{W}_{BA}$ is obtained by switching the order between $I_A$ and $I_B$. See Figure \ref{fig:stn_arch} for an illustration, and our supplementary material for specific design details.

\begin{figure}[h]
\adjincludegraphics[width=8.5cm,trim={{.1\width} 0 {.1\width} 0},clip]{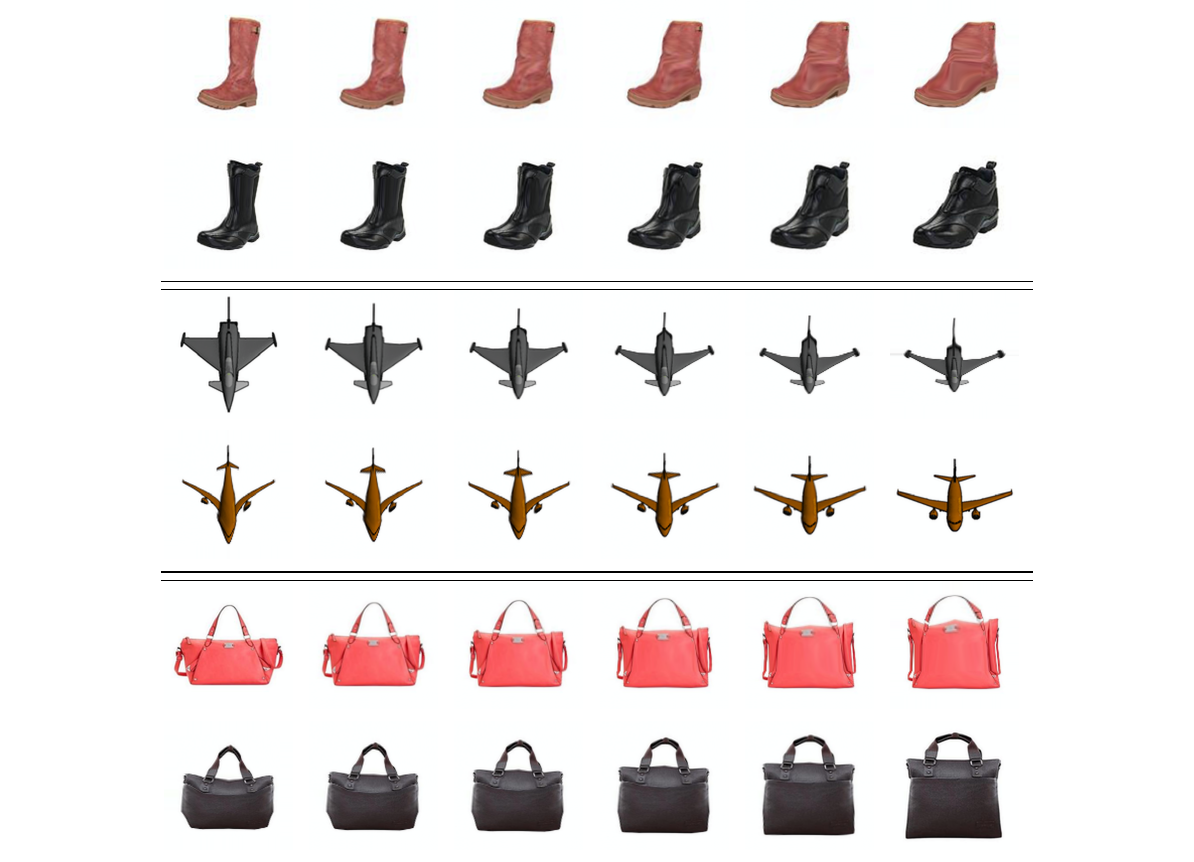}
\caption{STN warp examples. We show three examples for bidirectional warps computed by our STN. For each example, the first input is in the top row on the left and the second input is in the bottom row on the right. The STN computes a full warp, shown in the top row on the right for the first input and in the bottom row on the left for the second. The warped instances in between have all been deformed with corresponding partial warps.}
\label{fig:stn}
\end{figure}

We combine the STN with our sequence generation scheme, by applying a series of partial deformations to the inputs, each corresponding to a certain time stamp.
The partial deformation for $\mathcal{W}_{AB}$ at time $t$ is $\mathcal{W}_{AB}^t = \mathcal{I} + t \cdot (\mathcal{W}_{AB} - \mathcal{I})$, and $\mathcal{W}_{BA}^t = \mathcal{I} + (1-t) \cdot (\mathcal{W}_{BA} - \mathcal{I})$ for $\mathcal{W}_{BA}$, where $\mathcal{I}$ is the identity warp grid.
The grids are upsampled to the input image size using bilinear interpolation, and are applied onto $I_A$ and $I_B$ to obtain a sequence of warped inputs $\{I_A^t\}_{t=t_1}^{t_k}, \{I_B^t\}_{t=t_1}^{t_k}$, that are passed on to the encoder. See figure \ref{fig:stn} for three examples of partial to full deformations computed by our STN.

We add two losses tailored to our spatial transformer network. The first is a shape warp loss, $\mathcal{L}_W$, comparing the warped $I_A$, denoted by $I_A^{t_k}$, to $I_B$, and the second, $\mathcal{L}_I$, compares the predicted grid to the identity grid, for regularization. $\mathcal{L}_W$ makes another use of perceptual similarity by using the deep VGG features of layer group 5. These provide a higher level of abstraction that encourages the overall shape of the warped image to match the other endpoint, as opposed to stylistic details. The two losses are given by:

\begin{equation}
\begin{split}
    \mathcal{L}_W = \text{PS}_5(I_A^{t_k}, I_B) \\
    \mathcal{L}_I =  \text{MSE}(\mathcal{W_{AB}},\mathcal{I})
\end{split}
\end{equation}

We note that the losses we have described thus far, do not directly bind the inner frames to the inputs $I_A, I_B$. With the addition of the alignment computation, we are able to add a final perceptual similarity loss, $\mathcal{L}_E$, that draws a connection between each frame and its corresponding warped inputs, without restricting the ability of the frame to shift the shape of its underlying object. We choose layer group 4 for this purpose, to benefit from a combination of abstraction and a notion of finer detail, and compute a blend of similarities dependent upon the time stamp:

\begin{equation}
    \mathcal{L}_E = \sum_{i=1}^k (1-t_i) \cdot \text{PS}_4(I_{t_i}, I_A^{t_i}) + t_i \cdot \text{PS}_4(I_{t_i}, I_B^{t_i}) \\
\label{eq:global_ps}
\end{equation}

The total loss function of our generator is thus:

\begin{equation}
    \mathcal{L}_G = \lambda_G\mathcal{L}_G + \lambda_T\mathcal{L}_T + \lambda_R\mathcal{L}_R + \lambda_W\mathcal{L}_W + \lambda_I\mathcal{L}_I + \lambda_E\mathcal{L}_E
\end{equation}

\subsection{Network structure}
The architectures of $G$ and $D$ are similar to those of DiscoGAN \cite{kim2017learning}. 
$G$ is composed of an encoder containing blocks of \textit{conv} and \textit{ReLU}
followed by a decoder, containing blocks of \textit{tconv} (transposed convolution) and \textit{ReLU}.
Both local and global $D$ contain blocks of \textit{conv} and \textit{ReLU} with a final \textit{Sigmoid}. In both $G$ and $D$, the number of blocks depends on the input image resolution. For more details please refer to our supplementary material.

\begin{figure*}
\centering
\includegraphics[width=15cm]{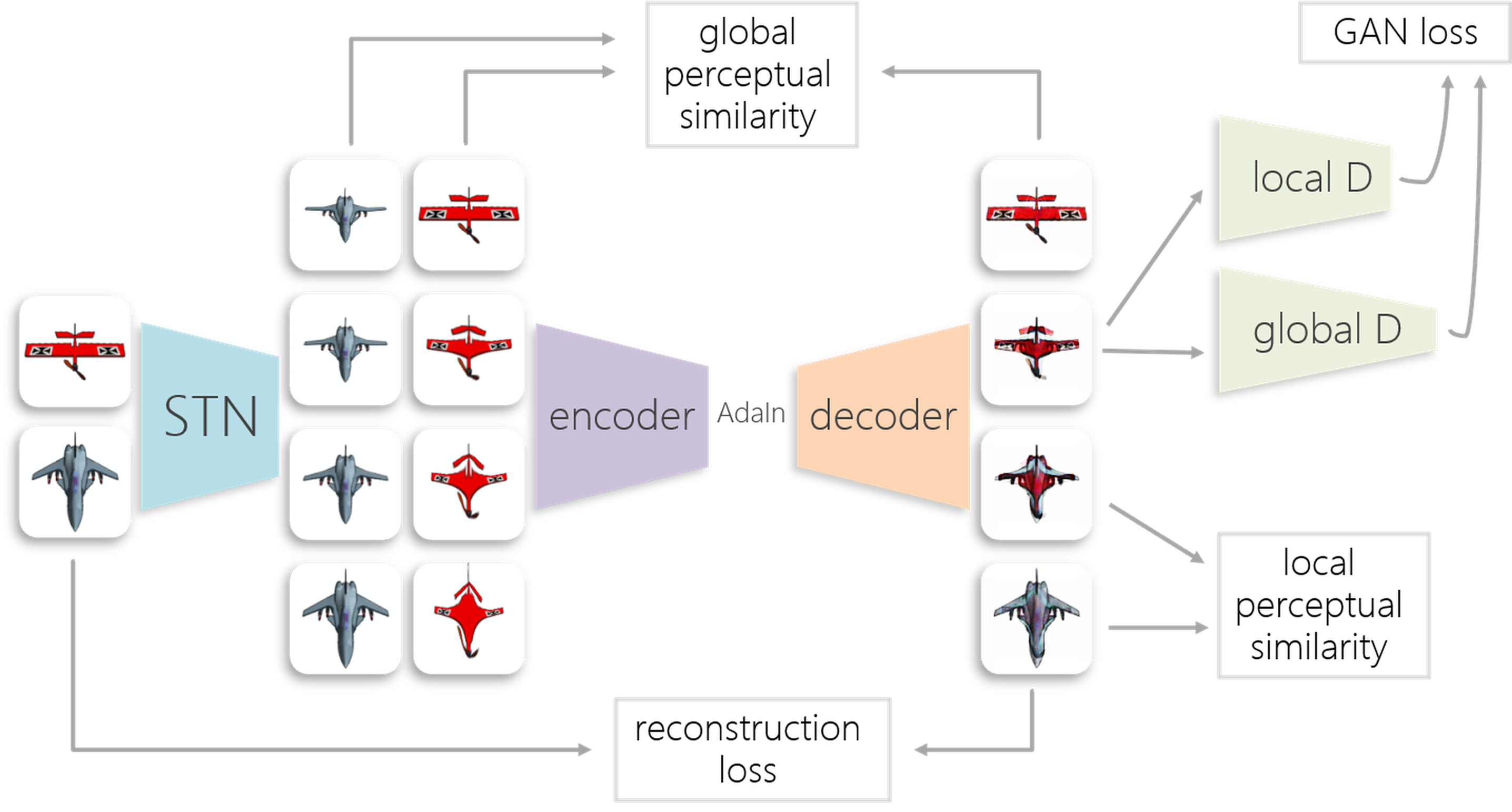}
\caption{Method pipeline. A pair of inputs is first processed by the STN. The predicted warp is applied onto the inputs to obtain a sequence of warped images corresponding to different time stamps. An encoder outputs a feature map per warped image, and every respective pair of feature maps undergoes a weighted adaptive instance normalization dependent on its time stamp, and proceeds, after channel-wise concatenation, into the decoder. All the resulting frames are evaluated by the GAN loss and a reconstruction loss compares the two endpoints to the inputs. Local and global perceptual similarity losses compare each pair of adjacent frames, and each frame to the warped inputs, respectively.  }
\label{fig:pipeline}
\end{figure*}

We employ a late fusion protocol, where the inputs $I_A^{t_i}, I_B^{t_i}$ are first processed separately by the encoder of $G$, which outputs feature maps $F_A^{t_i}, F_B^{t_i}$ respectively. An adaptive instance normalization component \cite{huang2017arbitrary} blends the mean and standard deviation of the feature maps according to the input time stamp $t_i$. That is, for given statistics $\mu_A^{t_i}, \mu_B^{t_i}$ and $\sigma_A^{t_i}, \sigma_B^{t_i}$, we compute the blended statistics for time $t_i$:

\begin{equation}
\begin{split}
    \mu_{t_i} = (1-t_i) \cdot \mu_A^{t_i} + t_i \cdot \mu_B^{t_i} \\
    \sigma_{t_i} = \sqrt{(1-t_i) \cdot (\sigma_A^{t_i})^2 + t_i \cdot (\sigma_B^{t_i})^2}
\end{split}
\label{eq:adain}
\end{equation}

$F_A^{t_i}$ is then updated as:
${F_A^{t_i}}^* = \sigma_{t_i} \cdot \frac{(F_A^{t_i} - \mu_A^{t_i})}{\sigma_A^{t_i}} + \mu_{t_i}$, and $F_B^{t_i}$ similarly.
Next, ${F_A^{t_i}}^*, {F_B^{t_i}}^*$ are concatenated channel-wise, along with an additional channel containing the time stamp $t_i$ expanded to the appropriate spatial resolution -- $F_{t_i}$. The resulting block of data, ${F_A^{t_i}}^* {F_B^{t_i}}^* F_{t_i}$, is processed by the decoder which outputs the corresponding generated frame. 
During training we generate $k$ frames, thus we prepare $k$ such blocks $\{{F_A^{t_i}}^* {F_B^{t_i}}^* F_{t_i}\}_{i=1}^k$, all of which are passed through the decoder. 

At train time, we randomly draw another instance from within the set for each input in the batch, and together these make up the input pairs. At each iteration, we also draw at random a pool of images to be shown to $D$ as \textit{real} data. Since each pair of inputs spawns $k$ frames, the \textit{real} pool for each pair is of size $k$ as well.
See Figure \ref{fig:pipeline} for a high level illustration of our pipeline.

\subsection{Content and style}
\label{sec:costl}
We extend our solution to address the problem of content and style separation \cite{gatys2015neural,johnson2016perceptual,dumoulin2017learned,huang2017arbitrary} within the morphing scope, to allow greater control over the desired outcome and provide increased freedom of creativity. Instead of a single axis of transformation between our two inputs $I_A, I_B$, we seek to engage two axes corresponding to disentangled transitions of content and style. This can be viewed as a 2D morphing effect taking place within the unit square, such that at coordinate $(t_{c_i},t_{s_j})$, the content of the generated frame reflects an interpolation of $(1-t_{c_i}) \cdot I_A^c + t_{c_i} \cdot I_B^c$ and its style a similar interpolation of $(1-t_{s_j}) \cdot I_A^s + t_{s_j} \cdot I_B^s$, where $t_{c_1},t_{s_1}=0$ and $t_{c_k},t_{s_k}=1$ ($k$ samples along both axes), and $I_A^c, I_B^c$ and $I_A^s, I_B^s$ are the content and style characteristics of the inputs respectively. 

We recognize the inherent capacity of the various components in our pipeline towards the distinction between the manifestation of content in our setup, \textit{i.e.}, overall shape and geometric detail, and stylistic attributes such as color and texture. Specifically, we observe that our local and global perceptual similarity losses can be employed in such a way as to encourage one aspect or the other by demand. Combining these with the initial warping mechanism catering to content (shape) rather than style, and the adaptive instance normalization component favoring style over content, we are able to formulate a disentangled solution dependent upon the two axes of transformation.

\textbf{Alignment.}
Initial alignment is carried out as before, but is only governed by the content axis, disregarding the style axis completely.

\textbf{Training.}
The new training protocol resembles our original one in that for each input pair, we generate $k$ frames. We randomly sample $k-2$ points along each axis, and keep $t_{c_1},t_{s_1}=0$ and $t_{c_k},t_{s_k}=1$. As the feature maps corresponding to frame $i$, $F_A^{t_{c_i}}, F_B^{t_{c_i}}$, exit the encoder, we perform adaptive instance normalization according to the style axis alone, such that $t_i$ in Equation \ref{eq:adain} is replaced with $t_{s_i}$. We then concatenate the \textit{two} samples associated with frame $i$ -- $t_{c_i}, t_{s_i}$, each expanded to the appropriate spatial resolution as before, to the normalized feature stack. The stack given to the decoder is thus: ${F_A^{t_{c_i}}}^* {F_B^{t_{c_i}}}^* F_{t_{c_i}} F_{t_{s_i}}$.

\textbf{Perceptual similarity losses.}
We create a hard separation between the authorities of the two PS losses with respect to content and style. The local PS loss $\mathcal{L}_T$ is assigned to the content whereas the global loss $\mathcal{L}_E$ is assigned to style. For $\mathcal{L}_T$, $t_i$ in Equation \ref{eq:local_ps} is replaced with $t_{c_i}$. Similarly, for $\mathcal{L}_E$, $t_i$ in Equation \ref{eq:global_ps} is replaced with $t_{s_i}$. Additionally, to increase the emphasis upon stylistic elements, we compute $\mathcal{L}_E$ with VGG layer group 3 instead of 4.


\section{Evaluation}
In this section we perform various experiments to evaluate our method,
both within its own scope (\ref{sec:ablation}), and externally (\ref{sec:res}). We experiment on four datasets - boots \cite{finegrained,semjitter} ($\sim10\text{k}$) and handbags \cite{zhu2016generative} ($\sim12\text{k}$), depicting real-world objects, and cars ($\sim7\text{k}$) and airplanes ($\sim4\text{k}$), featuring renders of objects from ShapeNet \cite{chang2015shapenet}.
For each dataset, we randomly draw 100 pairs of inputs from a separate test set upon which we conduct all our experiments. For each pair, we generate a sequence of 11 frames. Our model is trained on a 128x128 image resolution for 200 epochs, except for the variations trained for the ablation study, which were trained on a 64x64 resolution for computational efficiency.

\subsection{Ablation study}
\label{sec:ablation}
We explore the individual contributions of our various design components by conducting an ablation study. For this purpose, we train six variations of our network, aside from the proposed solution. 
Each variation excludes one component: GAN loss (adversary), local perceptual similarity, global perceptual similarity, reconstruction loss, adaptive instance normalization and STN (which also excludes global perceptual similarity, see Subsection \ref{sec:alignment}).

\begin{figure*}
\newcommand{\ablfig}{1.4}
\newcommand{\ablfigsp}{0.2} 
\setlength\tabcolsep{1pt} 
\centering
\includegraphics[width=16cm]{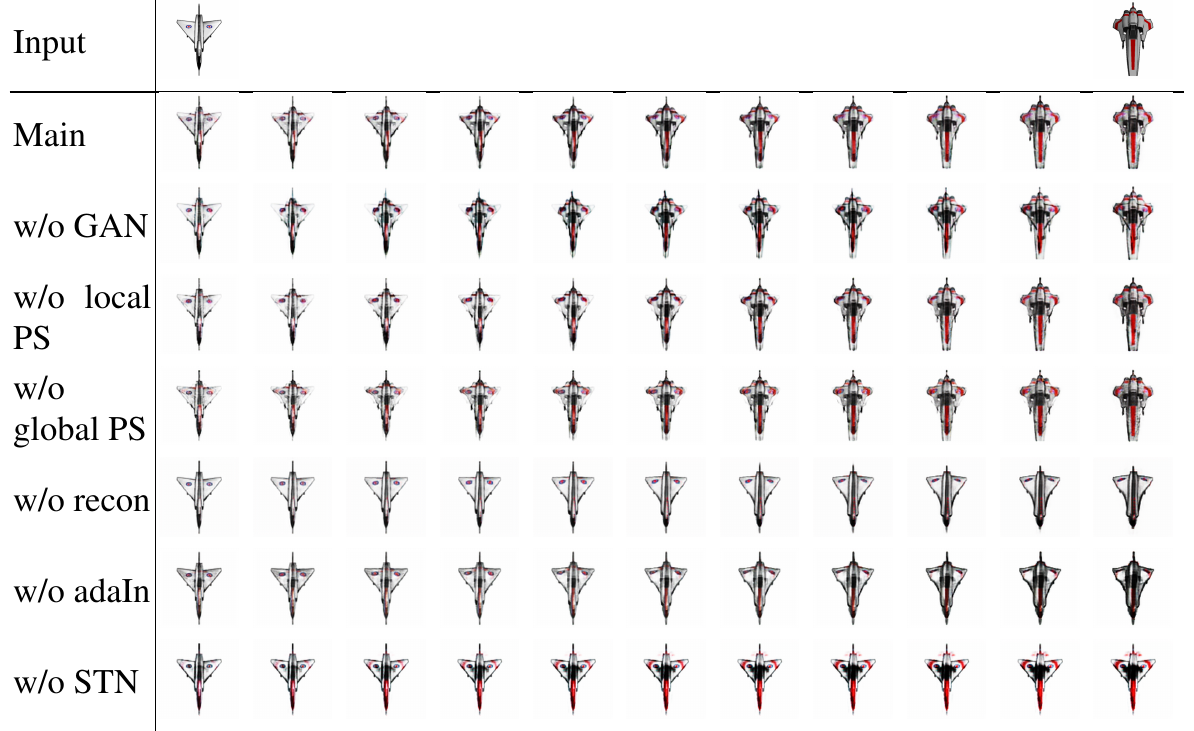}
\caption{Qualitative ablation. We present 11 frames generated for an input pair of planes2 by our main solution and the five ablation variations. All variations in this study operate on a resolution of 96x96, and were trained for 200 epochs.}
\label{fig:ablation}
\end{figure*}

We compute the Fr\'echet Inception Distance (FID) \cite{heusel2017gans} between the generated frames of each version in each dataset, and its respective training set, resized to a resolution of 96x96.
The overall trend of the scores, summarized in Table \ref{tbl:fid_ablation}, indicates that our main solution generates images that are generally in-line with the training set distribution.
Additionally, in Figure \ref{fig:ablation}, which contains visual examples for generated sequences obtained with the six variations, we note the various shortcomings characterizing the five ablation variations. The "w/o GAN" version does not preserve object detail, the "w/o PS" versions do not appropriately combine characteristics from both inputs, the "w/o recon" version does not adhere to the two endpoints and neither does the "w/o adaIn" version, 
and the "w/o STN" version is characterized by a serious degeneration, exhibiting little to no deformation in shape, resulting in a preference of one endpoint over the other. Note that as part of our earlier experiments, we did not experience a similar degeneration with a baseline system that did not incorporate an STN. However, these earlier versions naturally produced substantially lower quality results (due to their lack of advanced image alignment), and their far-removed architecture places them are outside the scope of this ablation study.

\begin{table}[h!]
\centering
 \begin{tabular}{| l | c c c c c|} 
 \hline
 Ablation & Bags & Boots & Cars & Planes & Mean \\ [0.5ex] 
 \hline
 Main & 31.96 & 27.75 &34.90 & 44.18 & 34.70 \\
 w/o GAN & 30.71 & 27.98 & 37.10 & 44.52 & 35.08 \\
 w/o local PS & 31.67 & 27.32 & 29.72 & 43.79 & 33.13 \\
 w/o global PS & 36.17 & 31.85 & 38.61 & 49.19 & 38.96 \\
 w/o recon & 33.18 & 29.03 & 36.13 & 41.17 & 34.88 \\
 w/o adaIn & 34.40 & 32.57 & 40.29 & 44.51 & 37.94 \\
 w/o STN & 53.68 & 57.72 & 64.18 & 57.26 & 58.21 \\
 
 \hline
 \end{tabular}
 \caption{Ablation FID scores on four datasets. We compute FID scores for different versions of our method, on a test set of 100 input pairs per dataset with 9 frames each, totaling at 900 frames per dataset. These generated frames are compared against the corresponding training set.}
\label{tbl:fid_ablation}
\end{table}

\begin{figure*}[h]
\newcommand{\cmpfig}{1.4}
\newcommand{\cmpfigsp}{0.2} 
\setlength\tabcolsep{1pt} 
\centering
\includegraphics[width=16cm]{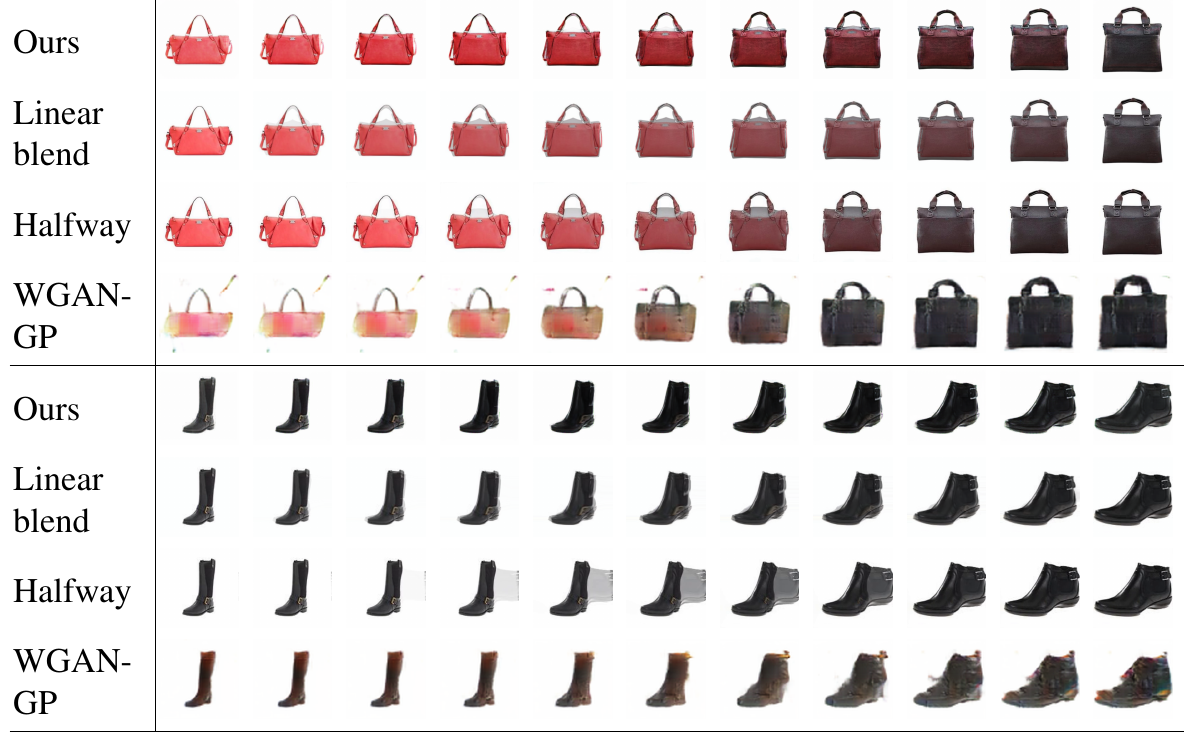}
\caption{Qualitative comparisons. Morphing sequences of bags and boots, generated by our method vs. three others.}
\label{fig:comparison2}
\end{figure*}

\begin{figure*}
\newcommand{\cmpfig}{1.4}
\newcommand{\cmpfigsp}{0.2} 
\setlength\tabcolsep{1pt} 
\centering
\includegraphics[width=16cm]{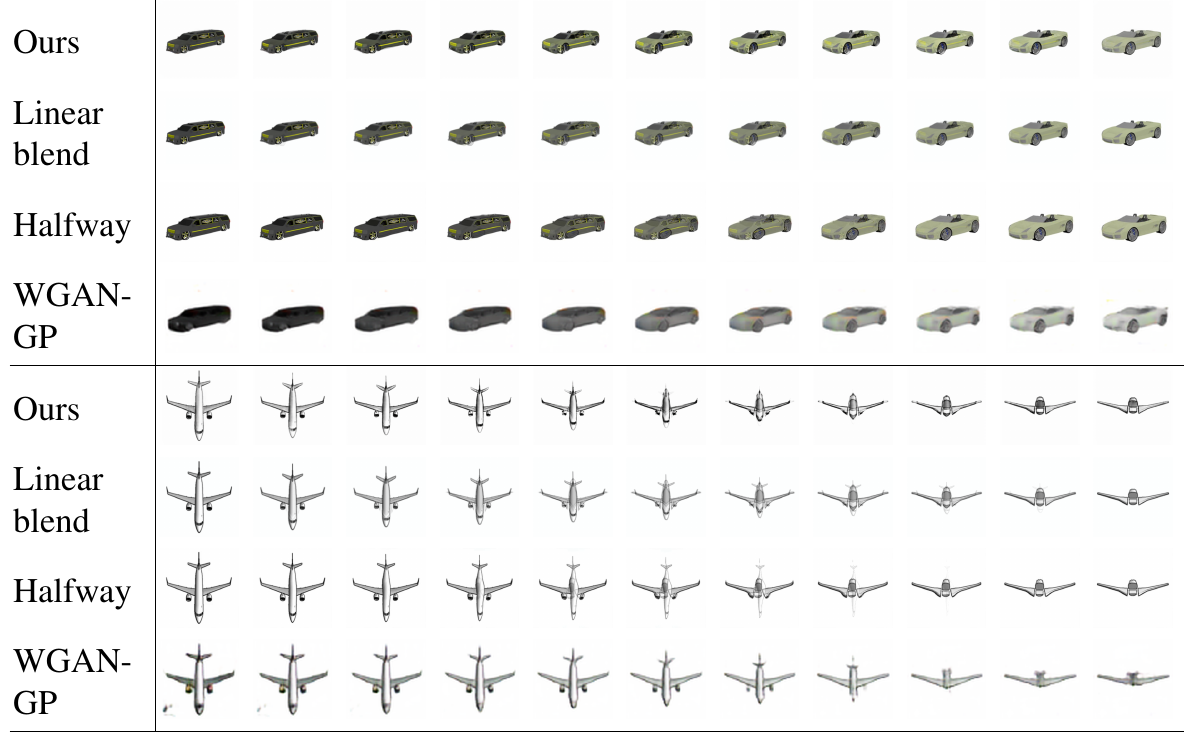}
\caption{Qualitative comparisons. Morphing sequences of cars and planes, generated by our method vs. three others.}
\label{fig:comparison3}
\end{figure*}

\subsection{Results and comparisons}
\label{sec:res}
We compare our results to three other methods.
The first is simple linear blending. We take the two sequences of warped inputs that our STN outputs, and blend each pair of corresponding frames according to their respective time stamp.
The second is the morphing method by Liao et al. \cite{liao2014automating} (termed "Halfway" in Table \ref{tbl:fid_comparison} and Figures \ref{fig:comparison2} and \ref{fig:comparison3}).
The final method is GAN-based latent space interpolation. 
Although recent high resolution GAN solutions such as BigGAN \cite{brock2018large} have been shown to produce impressively high quality generation and interpolation results, they are not as readily available to train, thus we opt for the well-known WGAN-GP \cite{gulrajani2017improved} solution for which we make use of the official implementation. We also experimented with VAE-GAN \cite{larsen2015autoencoding} and IntroVAE \cite{huang2018introvae}, but found WGAN-GP to provide superior results on our data. 
After training WGAN-GP on each of our four datasets, we train an encoder per trained model, to assist in our efforts to recover latent codes of existing instances. To obtain the latent codes of our test input images, we first pass them through the trained encoder, and then proceed to optimize the code further with an $L2$ loss on the input image. 

Table \ref{tbl:fid_comparison} summarizes the FID scores obtained by comparing the generated frames of each method on each of the test sets, with the corresponding training set. Note that all methods except WGAN-GP, which is compared at a 64x64 resolution, are compared at a resolution of 192x192.
The presented scores show that the classic techniques we compared to produce images that are closer to the ``real" data distribution than those generated by our method and WGAN-GP. These results are not surprising, since the classic techniques operate on the original images and perform operations of warp and cross dissolve, while generator-based methods procure the entire image every time, and are therefore bound to stray farther from the original distribution. Thus, even when an intermediate frame features ghosting artifacts, it may not incur a high FID score when it is essentially a blend of the two original inputs, as is the case in both of the classic methods.
				
\begin{table}[h!]
\centering
 \begin{tabular}{| l | c c c c c|} 
 \hline
 Comparison & Bags & Boots & Cars & Planes & Mean \\ [0.5ex] 
 \hline
 Ours & 29.12 & 25.78 & 28.94 & 50.35 & 33.55 \\
 Linear blend & 29.75 & 23.97 & 28.04 & 45.14 & 31.72 \\
 Halfway & \textbf{22.72} & \textbf{21.47} & \textbf{23.06} & \textbf{39.61} & \textbf{26.71} \\
 WGAN-GP & 68.91 & 83.71 & 54.38 & 55.53 & 65.63 \\
 \hline
 \end{tabular}
 \caption{Comparing FID scores on four datasets. 
 All methods were given the same set of 100 input pairs yielding morphing sequences of length 9, totaling at 900 frames per method. 
 }
\label{tbl:fid_comparison}
\end{table}

Figures \ref{fig:comparison2} and \ref{fig:comparison3} present qualitative examples of our generated sequences compared to those of the other methods. We observe that classic techniques exhibit excellent adherence to the original inputs as well as smooth transitions, however, at times they suffer from ghosting artifacts and exaggerated deformations due to incorrect correspondences. In contrast, our method is able to overcome differences in the overall shape, supporting a plausible transformation between the inputs.
Specifically, we note that Liao et al. \cite{liao2014automating} (Halfway) produce high quality effects composed of visually pleasing frames when the correspondence is accurate (many examples are available in our supplementary material). The difficulty arises when the two input images depict objects of larger shape offsets (see the Boots example in Figure \ref{fig:comparison2} and the Planes example in Figure \ref{fig:comparison3}), or are somewhat lacking in texture and color (see the wheels in the Cars example in Figure \ref{fig:comparison3}).
The baseline, linear blending, uses the alignment computed by our STN, and therefore benefits from its robustness to large differences in shape. However, the alignment provides general cues for warping, and further processing is often needed in order to promote smoother transitions. See Figure \ref{fig:comparison2}, where ghosting artifacts are visible just above the opening of the bag, and at the tip and back of the boots.
Lastly, our experiments with WGAN-GP \cite{gulrajani2017improved} show that generation quality as well as latent space encoding of existing instances, is still insufficient for high quality morphing effect creation. However, despite the artifacts that often appear in the generated frames, a strong advantage of latent space interpolation is its manner of frame creation. Frames are generated independently of one another, unlike approaches that are based on warp and cross-dissolve operations, and thus, ghosting artifacts are naturally avoided.

\subsubsection{User study}
To obtain user perspective, we designed a survey that presents the user with 36 pairs of morphing effects (9 of each dataset), such that each pair is composed of our result vs. that of one of the compared methods (in arbitrary order). For each pair, the users were asked to select the one they preferred of the two (subjectively), as well as the one that exhibits a more plausible transformation of shape (objectively). Users were able to select \textit{'no preference'} whenever they wished. A total of 50 participants took part in our study.
The results are shown in Table \ref{tbl:user_study}, where the column 'Ours' contains the portion of morphing effects where our method was selected over the other method (appearing in the 'Compared to' column). Likewise, the column 'Theirs' contains the portion where the other method triumphed, and the 'Tie' column specifies the remaining portion, where 'no preference' was selected. The statistics of the two questions appear in the same cell in the format $q_1/q_2$, such that $q_1$ corresponds to the statistics of the first question. These results show that in all sets except for Planes, users prefer Liao et al. \cite{liao2014automating} (Halfway) over ours, with larger margins in the real image datasets (Bags and Boots), where faithfulness to the original image statistics is more crucial. The Planes dataset contains instances with highly distinct silhouettes that prove challenging for all methods, but are slightly better handled by our method, which is able to reliably compute the alignment between the inputs. Our method had the upper hand over Linear blend and WGAN-GP in all datasets, with a smaller margin against Linear blend, whose performance is satisfactory when the two input shapes are sufficiently similar in shape, but otherwise produces ghosting artifacts. Note that all morphing effects were taken from the pool of 100 effects per dataset that we generated from the test set, all of which are available for viewing in our supplementary material.

While the classic method of Liao et al. \cite{liao2014automating} has the overall upper hand in terms of user preference, the advantage of our method is its consistency and robustness to different shape silhouettes and textures, and its speedy inference time (see our supplementary material for run time comparisons). Our main limitation is individual frame quality which relies on network generation, thus, latest and future advances in neural generation may help alleviate this, although at a probable training time penalty.

\begin{table}[h!]
\centering
 \begin{tabular}{| l l | l l l|} 
 \hline
 Set & Compared to & Ours & Theirs & Tie \\ 
 \hline
 Bags & Halfway & 0.25/0.27	& \textbf{0.625/0.59} & 0.125/0.13 \\
 Bags & Linear blend & \textbf{0.51/0.49} & 0.22/0.25 & 0.27/0.26 \\
 Bags & WGAN-GP & \textbf{0.88/0.875} & 0.08/0.08 & 0.04/0.046 \\
 \hline
 Boots & Halfway & 0.29/0.27 & \textbf{0.53/0.53} & 0.18/0.2 \\
 Boots & Linear blend &\textbf{ 0.39/0.41} & 0.32/0.28 & 0.29/0.3 \\
 Boots & WGAN-GP & \textbf{0.93/0.86} & 0.007/0.05 & 0.066/0.083 \\
 \hline
 Cars & Halfway & 0.34/0.33 & \textbf{0.45/0.45} & 0.21/0.23 \\
 Cars & Linear blend & \textbf{0.38/0.375} & 0.3/0.3 & 0.32/0.32 \\
 Cars & WGAN-GP & \textbf{0.89/0.88} & 0.01/0.04 & 0.09/0.08 \\
 \hline
 Planes & Halfway & \textbf{0.43/0.43} & 0.41/0.41 & 0.16/0.16 \\
 Planes & Linear blend & \textbf{0.47/0.48} & 0.3/0.3 & 0.23/0.21 \\
 Planes & WGAN-GP & \textbf{0.86/0.81} & 0.04/0.086 & 0.11/0.11 \\
\hline
 \end{tabular}
 \caption{User study results. Refer to main text for details.
 }
\label{tbl:user_study}
\end{table}

\begin{figure*}[t]
\newcommand{\csfig}{1.9}
\newcommand{\csfigg}{1.4}
\newcommand{\csfigsp}{0.2} 
\setlength\tabcolsep{1pt} 
\centering
\includegraphics[width=16cm]{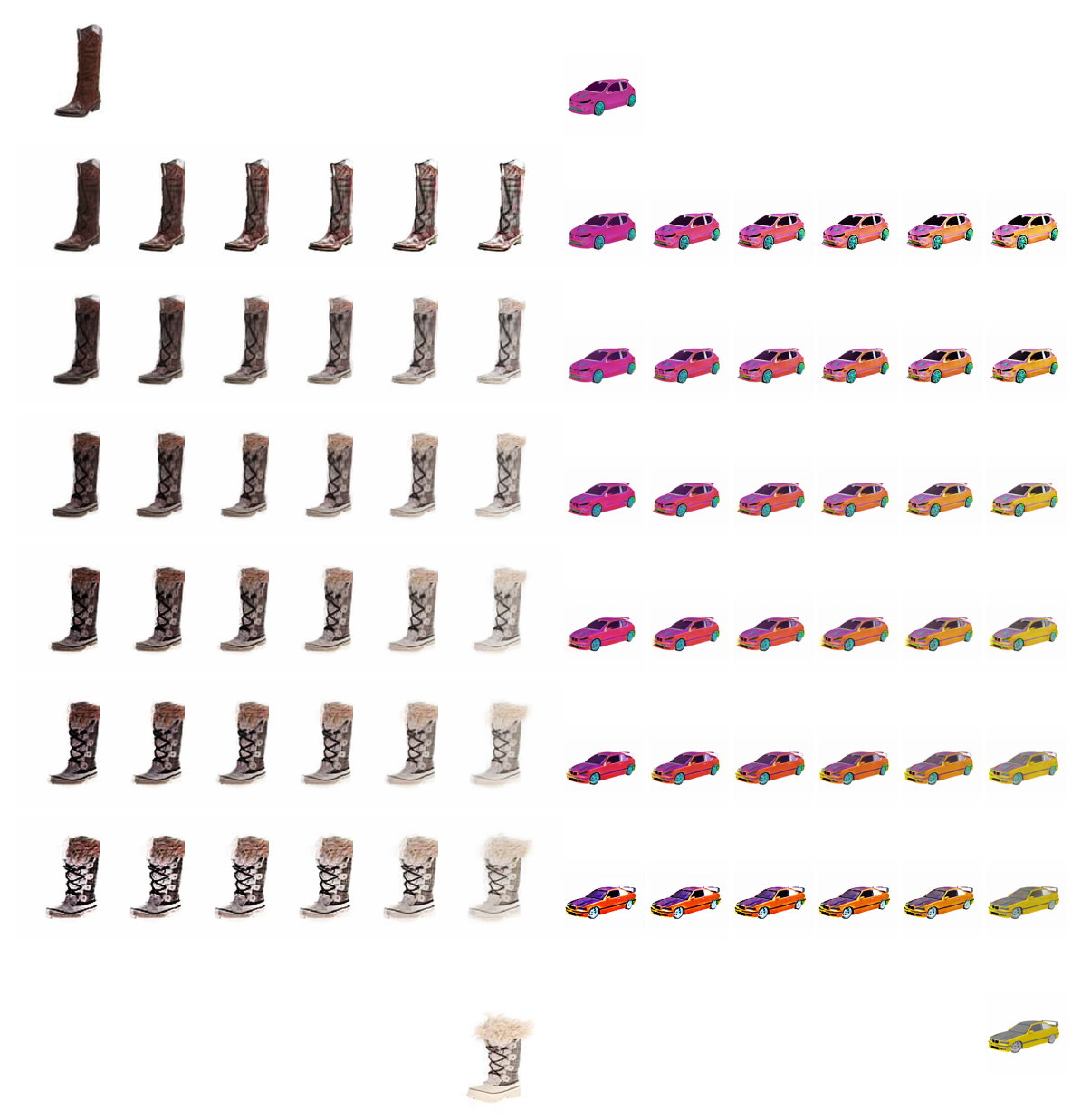}
\caption{Content and style disentangled morphing. A 2D 6x6 morphing grid between an input pair of boots (shown at the top left and bottom right corners) appears on the left, and similarly for cars on the right.}
\label{fig:costl}
\end{figure*}

\subsubsection{Content and style}
Figure \ref{fig:costl} contains two examples for content and style disentangled morphing as described in Subsection \ref{sec:costl}. For a given input pair, we generate each frame in a 6x6 grid, such that for cell $(i,j)$, the coordinate $i$ represents the desired location on the content axis, and similarly for coordinate $j$ with the style axis.

For more results, please see our supplementary material. For our full implementation please see our GitHub page.


\section{Conclusion}

We presented a new approach for morphing effect generation, combining the conditional GAN paradigm with a grid-based freeform deformation STN and a set of perceptual similarity losses. The components that make up our pipeline have been carefully curated to promote the generation of realistic in-betweens with smooth and gradual transitions, resulting in a solution that is robust to inputs exhibiting differences in shape and texture. Particularly, shape misalignments are overcome automatically by the integrated STN that learns a strong shape prior based on semantic features, rather than on potentially misleading low-level features.

In a world that is constantly hungry for more visual data, the ability to generate high-fidelity image instances is particularly beneficial. These can be used not only for artistic purposes, but also to enrich and augment existing datasets in support of various endeavors requiring substantial amounts of information.
Moreover, as a frame generation framework, a natural and potentially advantageous connection ties us to the field of video processing and synthesizing, one that may establish a bidirectional exchange of ideas with the prospect of mutual gain.

We note that our current setup is composed of simple building blocks -- a no-frills generator and discriminator that maintain a balance of good performance with low computational cost. Despite that,  
potential improvements and extensions to these components may further increase the quality of the generated frames, which are not always free of common morphing maladies such as ghosting and blurring. The addition of supervision to the pipeline may broaden the scope of our approach, and allow various types of transitions such as rotations. Similarly, morphing between images with arbitrary backgrounds may call for an integration of a dedicated segmentation component, one that is either pretrained, or trained within the entire framework in an end-to-end manner.

\section*{Acknowledgements}
This work was supported by Adobe and the Israel Science Foundation (grant no. 2366/16 and 2472/17).

\bibliographystyle{eg-alpha-doi}
\bibliography{egbib}

\end{document}